\documentclass[11pt]{article}
\usepackage[margin=1in]{geometry}
\usepackage{amsmath,amssymb,amsfonts,mathtools,bm}
\usepackage{amsthm}
\usepackage{booktabs,array,multirow}
\usepackage{graphicx}
\usepackage{microtype}
\usepackage[super,sort&compress]{natbib}
\usepackage[hidelinks]{hyperref}
\hypersetup{
  pdftitle={Memory compression and physical state augmentation favor different AMOC prediction tasks},
  pdfauthor={Mauricio Herrera-Marin},
  pdfsubject={AMOC prediction using physical-state augmentation and Mori--Zwanzig memory},
  pdfkeywords={AMOC, CMIP6, climate prediction, Mori--Zwanzig, memory, tipping risk}
}
\usepackage{caption,subcaption}
\usepackage{xcolor}
\usepackage{lineno}
\usepackage{siunitx}
\graphicspath{{figures/}}
\newtheorem{theorem}{Theorem}

\newcommand{\R}{\mathbb{R}}
\newcommand{\C}{\mathbb{C}}

\newcommand{\rad}{\operatorname{rad}}
\newcommand{\spec}{\operatorname{spec}}

\newcommand{\dd}{\mathrm{d}}
\title{Memory compression and physical state augmentation favor different AMOC prediction tasks}
\author{Mauricio Herrera-Mar\'in\thanks{Correspondence: \texttt{mherrera@udd.cl}; ORCID 0000-0002-9604-3077}\\
\small Faculty of Engineering, Universidad del Desarrollo, Santiago, Chile}
\date{}

\begin{document}
\maketitle

\begin{abstract}
The Atlantic Meridional Overturning Circulation is monitored and emulated through reduced indices, but such projections discard thermohaline structure and may require either explicit physical state or memory of the observed index. We compare these strategies in 30 branch-consistent CMIP6 trajectories from eight model families using leave-one-family-out validation. Salinity, temperature and density information improves direct 20-year forecasts, whereas compact scalar memory is top-ranked at every recursive horizon and yields the lowest case-averaged Brier score. A matched ablation confirms that feedback from memory improves long-horizon prediction. Physical state and recent trends also predict future ocean-state changes beyond the emissions pathway, most robustly at five years. NorESM under SSP5--8.5 identifies a forcing-dependent limit of scalar compression, while MIROC shows negative long-horizon transfer. A resolvent analysis explains why stable memory components do not guarantee stability of the complete learned model. Physical augmentation and memory compression therefore serve different AMOC prediction tasks.
\end{abstract}


\section*{Introduction}
The Atlantic Meridional Overturning Circulation (AMOC) redistributes heat, freshwater, carbon and nutrients through the Atlantic and links high-latitude water-mass transformation to climate far beyond the ocean basin. A substantial weakening would alter North Atlantic heat uptake, European climate, tropical rainfall, marine productivity and regional sea level. The assessed consensus is that the AMOC will very likely weaken during the twenty-first century under all Shared Socioeconomic Pathways, while the magnitude of weakening and the likelihood of an abrupt transition remain uncertain \cite{ipcc2021}. This uncertainty is not a peripheral detail: it limits the interpretation of early-warning indicators, the construction of reduced emulators for climate-risk assessment and the use of multi-model ensembles to estimate forced AMOC change.

The observational problem is intrinsically one of partial observation. Trans-basin arrays directly measure overturning and its heat and freshwater transports, but their records are short relative to multidecadal variability and forced adjustment \cite{buckley2016,frajka2019,weijer2019}. Longer reconstructions therefore rely on fingerprints such as subpolar sea-surface temperature, salinity, density gradients, air--sea heat fluxes or freshwater transport. These fingerprints have supported evidence for historical weakening and loss of stability \cite{rahmstorf2015,boers2021,ditlevsen2023,vanwesten2024}, but their interpretation is contested because internal variability, atmospheric forcing and non-stationary proxy--AMOC relationships can mimic or obscure a circulation signal \cite{benyami2024,latif2022,terhaar2025}. Recent observations and reconstructions continue to sharpen, rather than remove, this tension.

The model evidence is equally structured. Rare-event simulations and biased freshwater budgets show that noise, mean-state errors and multistability can reorganize transition pathways \cite{cini2024,boot2025}. At the same time, a 34-model analysis found a weakened but persistent overturning branch even under extreme greenhouse-gas and freshwater forcing \cite{baker2025}. Across CMIP6, the spread in future weakening is related to the simulated mean state and to the pathways by which deep and upper-ocean waters connect the Atlantic to the Southern Ocean and Indo-Pacific \cite{baker2023}; observational constraints can therefore shift projected weakening substantially \cite{portmann2026}. Western-boundary observations additionally indicate a meridionally coherent decline that is only partly compensated elsewhere in the basin \cite{xing2026}. Together, these studies show that AMOC uncertainty is not simply uncertainty in one scalar amplitude. It reflects how different models represent thermohaline structure, overturning pathways, forced drift and memory across timescales.

This creates a specific gap for reduced climate prediction. Most AMOC warning studies choose a scalar fingerprint and ask whether its trend, variance or autocorrelation changes. Most emulators instead enlarge the predictor set with physical covariates. What has not been established is whether discarded ocean structure should be retained as \emph{instantaneous physical state} or represented as \emph{memory of a reduced AMOC observable}, and whether the answer changes between two climate tasks: a forecast fitted separately at a fixed lead and a single learned operator propagated recursively. These tasks are often conflated, although they impose different requirements. Fixed-horizon prediction rewards any state information correlated with the target at that lead. Recursive rollout additionally demands a compact representation whose own forecast errors do not amplify under repeated application.

Mori--Zwanzig projection provides a principled framework for this comparison. Eliminating unresolved coordinates yields an instantaneous resolved tendency, a history-dependent feedback and orthogonal dynamics \cite{mori1965,zwanzig1973,chorin2013}. In climate dynamics, this formalism justifies delay and non-Markovian closures rather than adding lags only heuristically \cite{falkena2019,kondrashov2015,lin2021}. Recent reduced-order work likewise shows that memory can improve partially observed dynamics, but it also raises a stability question that is especially relevant for climate emulators: stable hidden memory modes do not necessarily imply that the complete resolved--memory map is contractive \cite{gupta2025,buitrago2025}.

\begin{figure}[t]
\centering
\includegraphics[width=\textwidth]{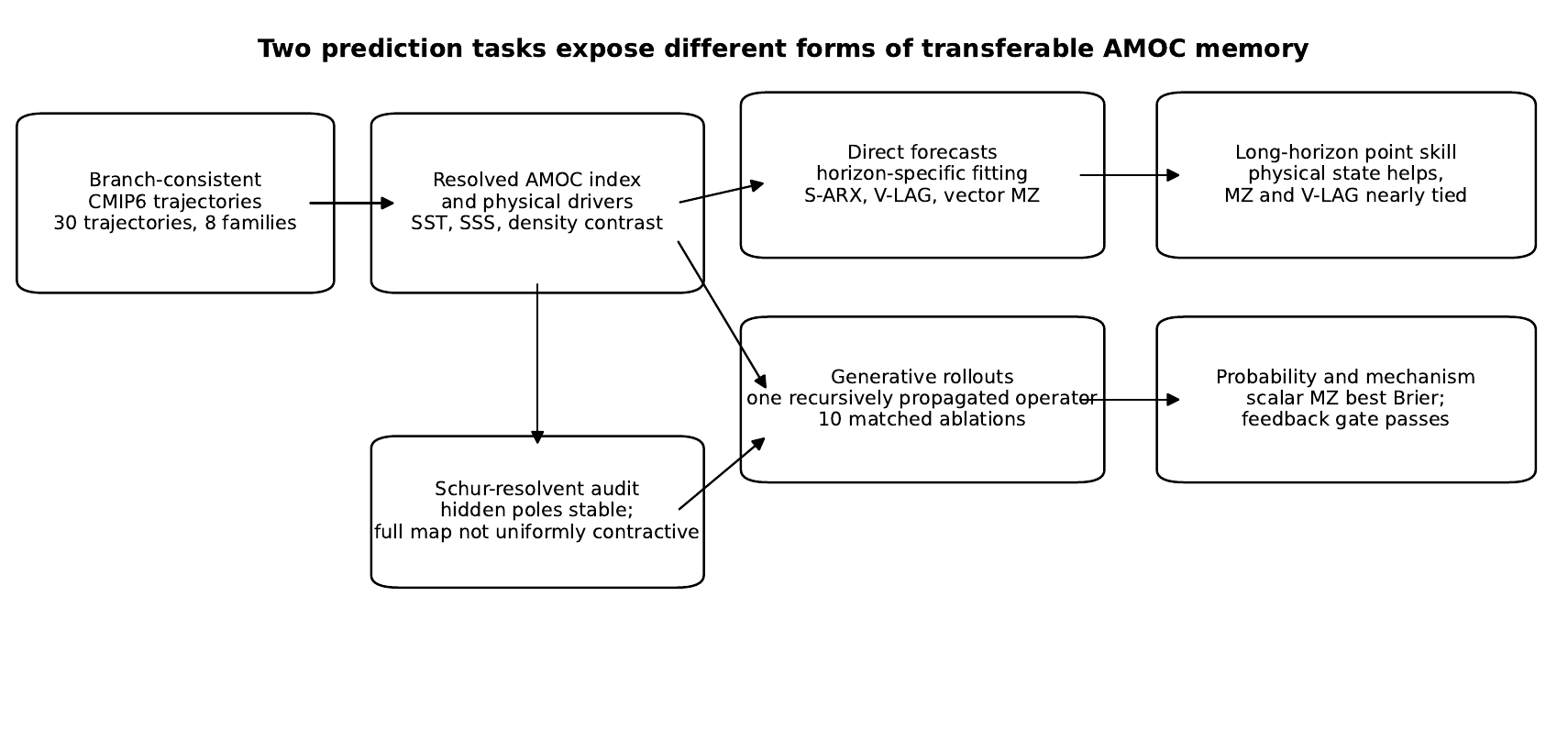}
\caption{\textbf{Cross-family design separates fixed-horizon and recursive AMOC prediction.} Branch-consistent CMIP6 trajectories support two complementary tasks. Horizon-specific direct forecasts measure the value of explicit physical state at a fixed lead, whereas recursive rollouts measure transfer of one propagated operator. Schur--resolvent diagnostics distinguish stable hidden memory from local stability of the complete resolved--memory map. All preprocessing and model selection exclude the held-out climate-model family, and no prediction uses future physical forcing.}
\label{fig:design}
\end{figure}

Here we connect this projection problem to cross-model AMOC prediction. We derive a matrix Schur--resolvent criterion that identifies the restoring balance after hidden variables are eliminated and separates hidden-pole stability from stability of the complete learned operator. We then test scalar memory, vector physical state and finite-lag alternatives on 30 branch-consistent historical--scenario trajectories from eight CMIP6 model families. The outer split leaves out an entire family, so success measures transfer across structural model differences rather than interpolation among members of the same family. We evaluate direct prediction, recursive rollout, probabilistic first-passage risk, matched memory ablations, cross-family physical-state information and local Jacobian geometry. Figure~\ref{fig:design} summarizes the design. Our principal finding is bidirectional: explicit thermohaline information improves fixed-horizon 20-year forecasts, whereas compact scalar memory is the most consistently top-ranked representation for recursive transfer. The forcing-dependent NorESM crossover and the negative MIROC transfer identify where that compression ceases to be sufficient.

\section*{Results}

\subsection*{Prediction task changes the preferred AMOC representation}
\begin{figure}[t]
\centering
\includegraphics[width=\textwidth]{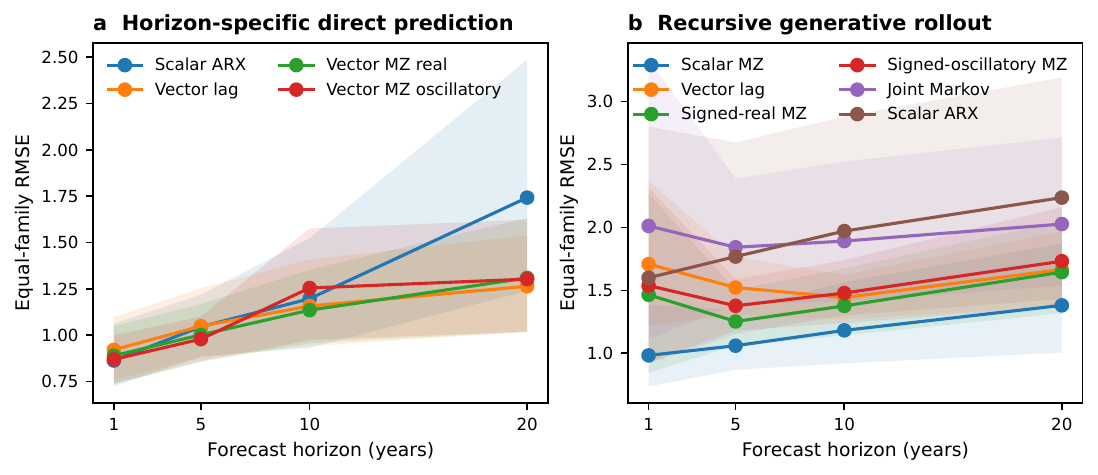}
\caption{\textbf{Fixed-horizon and recursive prediction favor different reduced states.} \textbf{a}, Equal-family RMSE for horizon-specific direct prediction; shading gives 95\% family-bootstrap intervals. Physical-state models improve the 20-year forecast, while vector lag and vector MZ are nearly tied. \textbf{b}, Equal-family RMSE for recursive rollout after averaging the three seeds within family; shading gives 95\% family-bootstrap intervals. Scalar MZ is top-ranked at every horizon.}
\label{fig:direct-recursive}
\end{figure}

The direct experiment fits an independent map at each lead. At one year, scalar ARX has the lowest pooled RMSE, consistent with a short-lead problem dominated by the recent AMOC state. At 20 years, the ordering reverses: scalar ARX has RMSE 1.595, whereas vector lag, oscillatory vector Mori--Zwanzig (MZ) and real vector MZ reach 1.297, 1.300 and 1.317, respectively. The 0.003 difference between vector lag and oscillatory MZ is negligible relative to family variation. Thus salinity, temperature and density summaries carry long-lead information that is absent from the scalar index, but no unique advantage of a vector MZ closure over an explicit finite-lag state is resolved (Fig.~\ref{fig:direct-recursive}a).

The recursive experiment asks a different question by propagating one learned operator over all horizons. Scalar MZ is top-ranked in mean error at 1, 5, 10 and 20 years (Fig.~\ref{fig:direct-recursive}b). At 20 years, mean family--seed MSE is 2.336 for scalar MZ, compared with 2.985 for vector lag, 3.042 for signed-real MZ and 3.355 for signed-oscillatory MZ. The corresponding equal-family RMSE is 1.381 for scalar MZ. The contrast between panels is the central empirical result: physical augmentation improves a lead-specific mapping, but the compact memory representation accumulates less cross-family error when its own outputs become future inputs.

The recursive ranking is consistent but not a claim of universal statistical superiority. Scalar MZ is better than vector lag in six of eight families at 20 years, yet the exact two-sided family test gives $p=0.3125$. None of the seven secondary scalar-MZ architecture comparisons survives Holm correction at 0.05; the minimum adjusted value is 0.0547. We therefore describe scalar MZ as the strongest default and the most consistently top-ranked recursive representation in this benchmark, not as uniformly superior to every vector alternative.

\subsection*{Forcing and model family expose structured limits of scalar compression}
\begin{figure}[t]
\centering
\includegraphics[width=\textwidth]{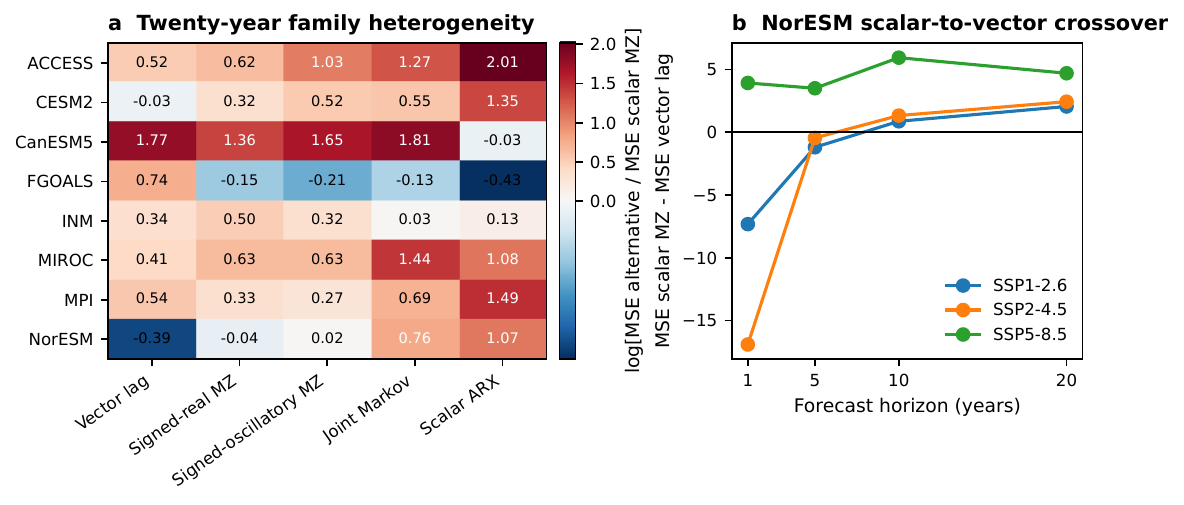}
\caption{\textbf{Recursive transfer is heterogeneous across climate-model families.} \textbf{a}, Log MSE ratio at 20 years for each alternative relative to scalar MZ; positive values favor scalar MZ. \textbf{b}, NorESM difference $\mathrm{MSE}_{\mathrm{scalar\,MZ}}-\mathrm{MSE}_{\mathrm{vector\,lag}}$. Positive values indicate a vector-lag advantage. The crossover occurs by 10 years in SSP1--2.6 and SSP2--4.5 and is present at every horizon in SSP5--8.5.}
\label{fig:heterogeneity}
\end{figure}

The family-level pattern reveals why a single average ranking is incomplete (Fig.~\ref{fig:heterogeneity}a). Scalar MZ has lower 20-year MSE than vector lag in ACCESS, CanESM5, FGOALS, INM, MIROC and MPI, is nearly tied in CESM2, and is worse in NorESM. The NorESM difference is organized by both lead and forcing pathway. Under SSP1--2.6 and SSP2--4.5, scalar MZ is better at 1--5 years but vector lag crosses over by 10 years. Under SSP5--8.5, vector lag is better from the first horizon and retains a 20-year MSE advantage of 4.70 (Fig.~\ref{fig:heterogeneity}b).

This crossover is not adequately described as generic out-of-distribution failure. NorESM origin states are well separated from the training-family state distribution, but other strongly separated families do not show the same ranking. Moreover, the NorESM conditional transition residual falls below the training-family reference at 20 years, when the vector-lag advantage is largest. The evidence instead points to a forcing-dependent limit of compression: recent multivariate physical history contains target-relevant structure that the scalar AMOC trajectory does not always retain.

\subsection*{Matched ablation identifies memory feedback rather than hidden capacity}
The confirmatory mechanistic comparison isolates the return path from hidden memory to the resolved AMOC increment. Signed-oscillatory MZ and its no-feedback ablation retain the same hidden coordinates and Schur-stable poles, but the ablation removes hidden-to-resolved feedback. At 20 years, the feedback model has lower MSE in seven of eight families, with mean paired difference $-2.280$ and exact two-sided $p=0.03125$. The result supports memory feedback as a predictive mechanism rather than hidden capacity alone. It does not establish a privileged oscillatory kernel: signed-real and signed-oscillatory variants are not resolved from each other, and vector lag remains statistically tied with the oscillatory model.

\subsection*{Probability accuracy and event discrimination favor different models}
Residual-resampled rollouts generate cumulative first-passage probabilities for a model-relative sustained AMOC weakening event. After averaging the three seeds for each unique forecast case, scalar MZ has the lowest Brier score at 5, 10 and 20 years. At 20 years its Brier score is 0.1441, ROC-AUC 0.8148 and average precision 0.6989. Mean forecast probability is 0.332 against prevalence 0.248, indicating moderate overprediction; the calibration slope is 0.906 and the intercept is $-0.599$.

\begin{figure}[t]
\centering
\includegraphics[width=\textwidth]{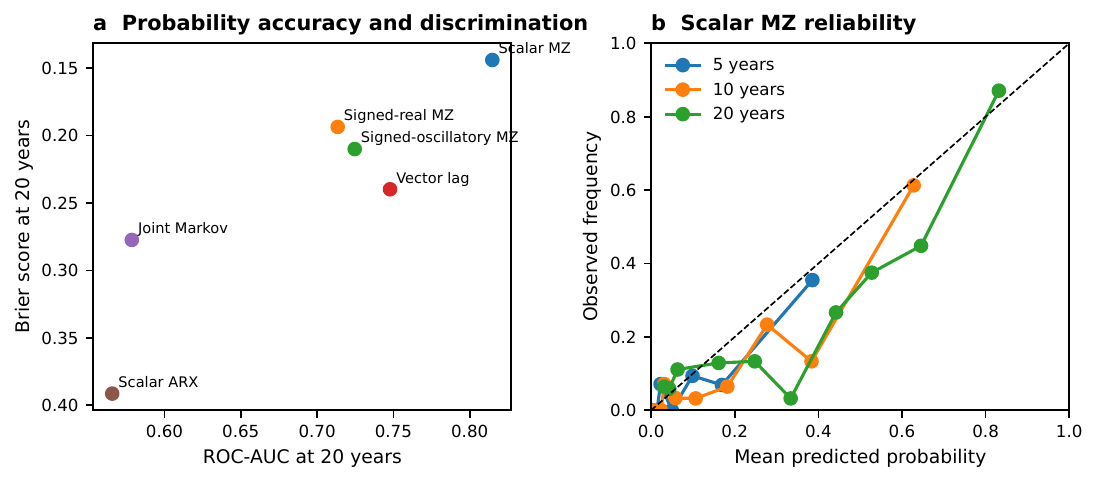}
\caption{\textbf{Probability accuracy and discrimination separate under long-horizon rollout.} \textbf{a}, Brier score versus ROC-AUC at 20 years after averaging seeds by forecast case. Lower Brier is better; scalar MZ is most accurate, whereas vector models retain stronger discrimination. \textbf{b}, Scalar-MZ reliability at 5, 10 and 20 years. The dashed line denotes perfect reliability.}
\label{fig:probability}
\end{figure}

The lowest Brier score does not imply the strongest event ranking. Vector lag and multivariate MZ models have higher macro-averaged discrimination at 20 years. Figure~\ref{fig:probability}a therefore separates probability accuracy from discrimination: scalar MZ produces the best average probabilities, whereas vector representations preserve stronger ranking information in some summaries. Reliability curves (Fig.~\ref{fig:probability}b) show close agreement at low risk and increasing long-horizon overprediction at intermediate probabilities. Scalar MZ has lower family-level Brier than four matched alternatives in all eight families, but none of nine secondary Brier comparisons survives Holm correction (minimum adjusted $p=0.0703$). The first-passage results are consequently used to compare representations, not to forecast an imminent real-world AMOC collapse.

\subsection*{Physical state retains dynamical information beyond the emissions pathway}
The four-dimensional origin state---subpolar salinity, subpolar and subtropical temperature, and their density contrast---contains strong model-family fingerprints. Across 22 available family--scenario combinations, mean domain-classifier AUC is 0.938 and held-out states frequently lie beyond the training-family support. Yet marginal state separation does not predict relative 20-year performance, and neither static Mahalanobis distance nor maximum mean discrepancy explains the NorESM crossover (Supplementary Note~8).

\begin{figure}[t]
\centering
\includegraphics[width=\textwidth]{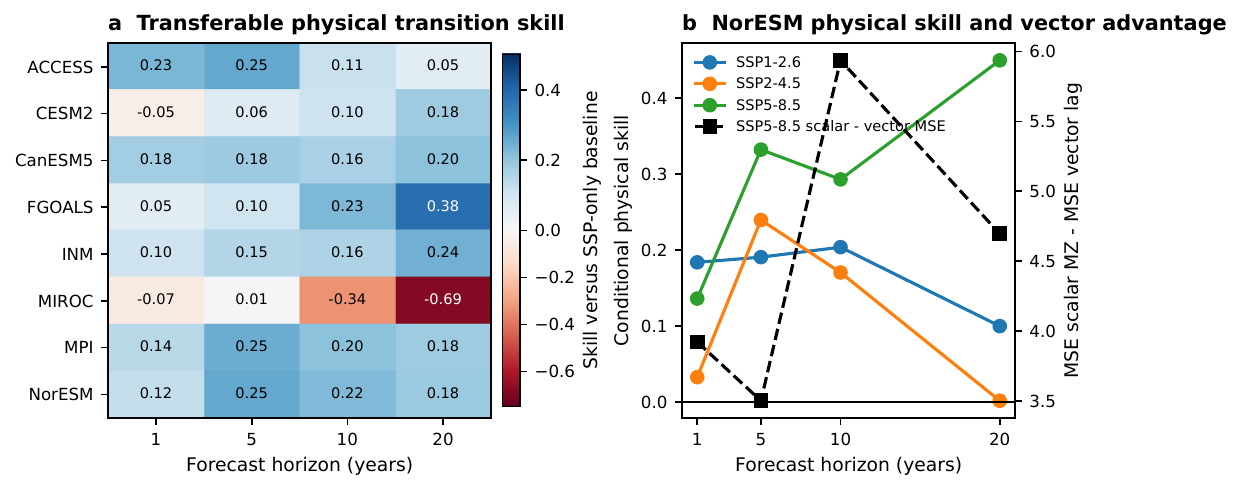}
\caption{\textbf{Thermohaline state carries information beyond the SSP pathway.} \textbf{a}, Conditional skill for predicting future physical-state changes relative to an SSP-only transition-mean baseline. Each cell averages the available scenarios within a family; FGOALS contributes SSP5--8.5 only. \textbf{b}, NorESM conditional skill by SSP and the SSP5--8.5 scalar-MZ minus vector-lag MSE difference. High physical skill and vector-lag advantage coexist under strong forcing, but the association is not systematic across families.}
\label{fig:physical-information}
\end{figure}

We therefore tested a more climate-relevant question: after conditioning on SSP, does present thermohaline structure contain transferable information about future ocean-state adjustment? For each horizon, a nested leave-one-family-out ridge model predicts $\Delta X_h=X_{t+h}-X_t$ from the origin state, retrospective 5- and 10-year trends and SSP indicators. The baseline is the training-family mean transition within the same SSP. Conditional skill is
\begin{equation}
\mathcal S_h=1-\frac{\mathrm{MSE}_{\mathrm{conditional}}}{\mathrm{MSE}_{\mathrm{SSP\ mean}}}.
\end{equation}
Mean equal-family skill is 0.088 at one year and 0.157 at five years, with 95\% family-bootstrap intervals $[0.020,0.152]$ and $[0.096,0.215]$. Five-year skill is positive in all eight families and remains significant in an exact two-sided family test after Holm correction across horizons ($p_{\mathrm{Holm}}=0.03125$). At 10 and 20 years, seven of eight families remain positive, but the intervals include zero because MIROC exhibits strong negative transfer (Fig.~\ref{fig:physical-information}a).

The long-horizon signal is strongest under SSP5--8.5. For NorESM, conditional physical skill reaches 0.332, 0.293 and 0.449 at 5, 10 and 20 years, respectively, while vector lag outperforms scalar MZ throughout that pathway (Fig.~\ref{fig:physical-information}b). Across the seven families represented in all three SSPs, the 20-year contrast between SSP5--8.5 and the mean of SSP1--2.6/SSP2--4.5 is positive in six families, but does not survive adjustment across horizons. We therefore interpret stronger long-horizon physical information under intense forcing as structured exploratory evidence. The general result is firmer: the physical state is not redundant with the emissions pathway, although its value for choosing an AMOC architecture is family dependent.

\subsection*{A resolvent criterion separates projected restoring feedback from full-map stability}
To connect the empirical rankings to reduced climate dynamics, consider a sampled lifted state with resolved coordinate $q_n\in\R^d$ and hidden memory $z_n\in\R^m$. Around a frozen state,
\begin{equation}
\begin{pmatrix}q_{n+1}\\z_{n+1}\end{pmatrix}
=J\begin{pmatrix}q_n\\z_n\end{pmatrix},\qquad
J=\begin{pmatrix}A&B\\C&D\end{pmatrix},\qquad \rad(D)<1.
\label{eq:block-map}
\end{equation}
Eliminating the hidden state generates the matrix lag kernel $K_k=BD^{k-1}C$.

\begin{theorem}[Schur--resolvent criterion]
For every $\zeta\notin\spec(D)$, define
\begin{equation}
\Phi(\zeta)=\zeta I_d-A-B(\zeta I_m-D)^{-1}C.
\end{equation}
Then
\begin{equation}
\det(\zeta I_{d+m}-J)=\det(\zeta I_m-D)\det\Phi(\zeta).
\label{eq:factorization}
\end{equation}
Writing $\mathcal N_j=B(I-D)^{-(j+1)}C$, a stationary unit multiplier occurs if and only if
\begin{equation}
\det(I_d-A-\mathcal N_0)=0.
\label{eq:unit-balance}
\end{equation}
\end{theorem}

Equation~\eqref{eq:unit-balance} is the exact restoring balance seen by the projected AMOC state. The instantaneous resolved block $A$ is renormalized by the integrated memory feedback $\mathcal N_0$; neither term is interpretable in isolation. For a simple slow direction, the same resolvent reconstructs the multiplier as
\begin{equation}
\rho_{\mathrm{slow}}=1+\frac{\delta}{\gamma}
+\frac{\ell^\top(\mathcal N_2+HGQH)r}{\gamma^3}\delta^2+O(\delta^3),
\label{eq:slow-multiplier}
\end{equation}
where $H=I_d+\mathcal N_1$, $Q=I-r\ell^\top$ and $G$ is the reduced inverse on the complement of the slow mode. The $HGQH$ term records coupling between the slow AMOC-like direction and the remaining resolved physical modes. Full proofs, the continuous-time limit, general unit-circle crossings, and stable signed and oscillatory realizations are given in Supplementary Notes~1--6.

\begin{figure}[t]
\centering
\includegraphics[width=\textwidth]{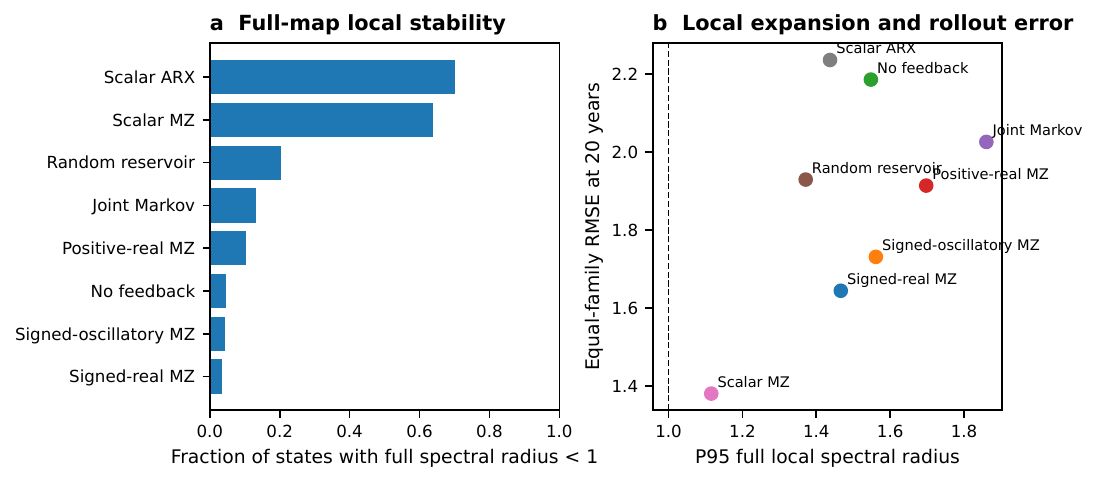}
\caption{\textbf{Stable memory components do not guarantee a contractive AMOC emulator.} \textbf{a}, Fraction of evaluated states with complete-Jacobian spectral radius below one. \textbf{b}, 95th-percentile full local spectral radius versus equal-family 20-year RMSE. The dashed line marks unit spectral radius. Hidden memory blocks are stable for all MZ fits, but complete maps are only intermittently contractive.}
\label{fig:stability}
\end{figure}

All fitted MZ realizations have Schur-stable hidden blocks, and all nonlinear rollouts remain finite. The complete local Jacobian is nevertheless not uniformly contractive. Scalar MZ is locally Schur stable at 64.0\% of evaluated states, compared with 3.5\% and 4.4\% for signed-real and signed-oscillatory MZ. Its 95th-percentile full spectral radius is also smaller (1.115 versus 1.466 and 1.561), and lower expansion co-occurs with lower 20-year error (Fig.~\ref{fig:stability}). Stable hidden memory therefore licenses the realization but does not establish stability of the complete learned climate emulator. Our claim is bounded predictive rollout with audited local geometry, not global asymptotic stability or evidence that the real AMOC is near a mathematical bifurcation.

\section*{Discussion}
The central climate result is that physical state augmentation and memory compression answer different prediction questions. Thermohaline structure improves direct 20-year forecasts because salinity, temperature and density gradients contain information about the future forced state that is absent from a single AMOC index. Recursive rollout imposes a second requirement: the representation must remain stable and transferable when forecast states are repeatedly reused. Under that requirement, scalar memory is the most consistently top-ranked default across model families and horizons. The two findings are complementary rather than contradictory.

This distinction matters for the interpretation of AMOC fingerprints. A reduced SST or salinity index is not expected to be a sufficient physical state; its apparent persistence can arise from unresolved ocean adjustment, remote forcing and coupled feedbacks. The present results show that part of this discarded information can be recovered through memory of the index, but not in every forcing regime. Consequently, early-warning and attribution studies should distinguish three propositions that are often merged: a fingerprint may correlate with AMOC strength, its history may improve prediction, and it may contain enough state information for recursive extrapolation. Our experiment supports the second proposition broadly, the third only conditionally, and does not use either as a direct estimate of real-world tipping time.

The results also inform climate emulators and multi-model constraints. A lead-specific emulator intended to estimate AMOC weakening at 2100 can benefit from explicit thermohaline predictors, consistent with the growing use of salinity, temperature and overturning-pathway information to constrain CMIP6 projections. A generative emulator used for first-passage probabilities, scenario exploration or sequential data assimilation should additionally control recursive error and full-map expansion. In such applications, selecting the state only by one-step or fixed-horizon skill can favor a representation that is less robust when propagated. Task-specific validation should therefore be treated as part of the climate question, not as a technical afterthought.

NorESM SSP5--8.5 and MIROC make the implications concrete. NorESM shows that strong forced drift can preserve long-horizon physical information and favor explicit vector lags even when scalar memory remains competitive elsewhere. MIROC shows the opposite failure mode: adding physical state can transfer negatively across families at long horizons. These exceptions argue against a universal AMOC fingerprint or a universal reduced state. They instead motivate adaptive emulators that use compact scalar memory as a cross-family baseline and activate multivariate physical history when trajectory diagnostics indicate forcing-dependent residual information.

The Schur--resolvent result provides the dynamical interpretation. Projecting a high-dimensional circulation does not remove unresolved feedback; it moves that feedback into a memory kernel. The relevant local threshold is therefore a Schur complement involving both resolved drift and integrated memory. Stable hidden poles are useful because they prevent an internally divergent memory realization, but they do not imply that the complete emulator is contractive. This matters for early-warning language: a learned spectral radius, autocorrelation increase or slow multiplier is a property of a chosen projection and fitted map. It should not be equated automatically with the stability margin of the full ocean circulation.

The statistical evidence is deliberately tiered. The matched feedback ablation is confirmatory and supports hidden-to-resolved memory feedback. Five-year conditional physical skill is also robust after correction across horizons. By contrast, scalar MZ versus vector lag is a descriptive ranking supported by consistency across recursive horizons and Brier scores, not by a significant pairwise test with eight families. This distinction limits the headline to what the design can establish: scalar memory is the strongest default in the evaluated cross-family rollout benchmark, while physical augmentation remains indispensable for fixed-lead prediction and identifiable forced regimes.

Several boundaries remain. Eight model families provide strict structural validation but limited family-level power, and shared components reduce their effective independence. FGOALS contributes SSP5--8.5 only, so balanced-family sensitivity is necessary for scenario averages. Surface salinity, temperature and density summaries do not exhaust freshwater transports, deep-ocean heat content, wind-driven pathways or Southern Ocean controls. The first-passage threshold is model relative and is not an estimate of collapse probability in the observed climate. Current observationally constrained projections and reconstructions remain in tension \cite{terhaar2025,portmann2026,xing2026}; this study addresses how reduced representations transfer across climate models rather than adjudicating those observational claims.

The next step is to couple representation selection to observations and process diagnostics. Direct arrays, boundary-density estimates, freshwater transports and emerging subsurface fingerprints could be used to determine when a scalar history is sufficient and when additional physical coordinates are required. Stability-constrained training of the complete Jacobian, rather than only the hidden block, offers a complementary route. The broader implication is that uncertainty in AMOC prediction is partly uncertainty about representation: what appears as a loss of predictability in one reduced index may be recoverable as memory, while what appears as memory failure may signal missing thermohaline state.

\section*{Methods}

\subsection*{Continuous and sampled hidden-variable elimination}
For a continuous linearization with resolved state $q\in\R^d$ and hidden state $s\in\R^m$,
\begin{equation}
\frac{\dd}{\dd t}\begin{pmatrix}q\\s\end{pmatrix}
=\begin{pmatrix}A_c&B_c\\C_c&D_c\end{pmatrix}\begin{pmatrix}q\\s\end{pmatrix},
\end{equation}
a Hurwitz $D_c$ gives the exact Volterra equation
\begin{equation}
\dot q(t)=A_cq(t)+\int_0^t B_ce^{D_c(t-\tau)}C_cq(\tau)\,\dd\tau+B_ce^{D_ct}s(0).
\end{equation}
The integrated feedback is $\mathcal X_0=-B_cD_c^{-1}C_c$. In sampled time, eliminating $z_n$ from Eq.~\eqref{eq:block-map} gives $K_k=BD^{k-1}C$. Resolvent moments $\mathcal N_j=B(I-D)^{-(j+1)}C$ yield Eqs.~\eqref{eq:unit-balance} and \eqref{eq:slow-multiplier}. Proofs, the continuous--discrete limit and the general unit-circle crossing formula are supplied in the Supplementary Information.

\subsection*{CMIP6 cohort and climate variables}
The study uses the Coupled Model Intercomparison Project Phase 6 archive \cite{eyring2016}. The audited cohort contains 30 branch-consistent historical--scenario trajectories from eight model families: ACCESS, CESM2, CanESM5, FGOALS, INM, MIROC, MPI and NorESM. Ten trajectories follow SSP1--2.6, ten SSP2--4.5 and ten SSP5--8.5. Historical and scenario segments are stitched only when source, member, grid and variable identity agree. The physical source variables are sea-surface salinity (\texttt{sos}), sea-surface temperature (\texttt{tos}) and grid-cell area (\texttt{areacello}). Annual summaries include subpolar salinity, subpolar and subtropical temperature, density proxies and the subpolar--subtropical density contrast. The resolved target is an annual AMOC-strength index. The database passed duplicate-key, annual-coverage and historical-to-scenario transition audits.

All standardization, learned feature transformations and hyperparameter selection are fitted within the outer training families. The held-out family is excluded from preprocessing and selection. Models receive only physical values available at the forecast origin and retrospective information. No model receives future physical forcing. This design tests transfer across climate-model structure rather than interpolation among trajectories from the same family.

\subsection*{Direct horizon-specific prediction}
A separate predictor is fitted at each horizon $h\in\{1,5,10,20\}$ years. Models are scalar ARX, vector lag, signed-real vector MZ and damped-oscillatory vector MZ. Hyperparameters are selected using training families only. Performance is reported as RMSE, with family-bootstrap intervals based on the eight held-out families. This protocol tests representation value at a fixed lead and is not interpreted as a recursive dynamical simulation.

\subsection*{Recursive rollout and matched ablations}
The recursive operator propagates resolved and hidden states as
\begin{align}
q_{n+1}&=8\tanh\!\left(\frac{q_n+\Delta q_\theta(q_n,z_n,u_n)}{8}\right),\\
z_{n+1}&=Cq_n+Dz_n.
\end{align}
The saturation lies outside the observed standardized range and prevents non-finite trajectories; it is part of the propagated map rather than a post-hoc clip. The ten architectures are scalar MZ, vector lag, signed-real MZ, signed-oscillatory MZ, positive-real MZ, shifted-memory placebo, random-reservoir memory, no-feedback memory, capacity-matched joint Markov and capacity-matched scalar ARX. The final design contains eight held-out families, ten architectures and three random seeds, for 240 jobs. All 240 completed, with no failed or missing jobs.

The prespecified mechanistic comparison is signed-oscillatory MZ versus its no-feedback ablation at 20 years. The broader scalar-MZ architecture comparisons are secondary. Seeds are averaged within family before family-level inference.

\subsection*{First-passage probability audit}
The event is the first occurrence of at least three consecutive annual AMOC values below 70\% of the model-specific historical mean, conditional on no previous event. Residual-resampled stochastic rollouts yield cumulative event probabilities. Probabilities from the three seeds are averaged for each unique trajectory--origin--horizon case before scoring, leaving 306 cases per model and horizon. Metrics are Brier score, log loss, ROC-AUC, average precision, calibration intercept, calibration slope and the reliability--resolution--uncertainty decomposition \cite{brier1950,murphy1973}.

\subsection*{Cross-family state shift and conditional physical dynamics}
Static shift is evaluated in the four-dimensional origin state $(\mathrm{sos}_{SPG},\mathrm{tos}_{SPG},\mathrm{tos}_{STG},\Delta\rho_{SPG-STG})$ using Ledoit--Wolf Mahalanobis distance, training-support exceedance, nearest-neighbor distance, radial-basis maximum mean discrepancy and a held-family domain classifier. Because annual origins within a trajectory are dependent, these statistics are treated primarily as descriptive diagnostics.

For the conditional transition analysis, the target at horizon $h$ is $\Delta X_h=X_{t+h}-X_t$. Predictors are $X_t$, five- and ten-year retrospective trends and SSP indicators. A multi-output ridge model is trained on seven families and evaluated on the eighth. The ridge penalty is selected by an inner leave-one-family-out loop. The baseline is the training-family mean $\Delta X_h$ within SSP. Residual normalization uses scenario-specific inner out-of-family predictions rather than in-sample errors. Family-cluster bootstrap intervals use 5,000 replicates.

\subsection*{Local stability diagnostics}
Automatic differentiation of the complete propagated map gives local Jacobian blocks $A,B,C,D$. Hidden stability is $\rad(D)<1$. Full local stability is $\rad(J)<1$ for $J$ in Eq.~\eqref{eq:block-map}. We report the stable fraction, 95th percentile and maximum over evaluated states. Hidden-pole stability is checked separately because it does not imply full-map contraction.

\subsection*{Statistics and reproducibility}
Climate-model family is the inferential unit ($n=8$). Exact two-sided sign-flip tests enumerate all $2^8$ family sign configurations and require no normality assumption. The prespecified feedback contrast is evaluated separately at $\alpha=0.05$. Secondary architecture families are corrected by Holm's sequential procedure \cite{holm1979}. Family-bootstrap intervals resample families with replacement and preserve all cases within the selected family. For conditional transition skill, two-sided exact tests across the four horizons are Holm-adjusted; only the five-year effect remains below 0.05. Cross-family shift--performance associations are exploratory and none survives correction across the tested association family. Exact sample sizes, comparison families and unadjusted and adjusted $p$ values are reported in the Supplementary Information and source tables.

\subsection*{Code availability}
The analysis code, frozen protocol, environment records, processed source data supporting all figures and tables, derived results, and reproducibility audits are available at \url{https://github.com/mauricio-herrera/amoc-mz-physical-v4}. The version-specific public package is archived on Zenodo at \url{https://doi.org/10.5281/zenodo.21606982}. Raw CMIP6 files are not redistributed; their provenance, model and member identifiers, processing contracts, validation audits, and reconstruction procedures are documented in the archived release.

\section*{Data availability}
CMIP6 source fields are available through the Earth System Grid Federation under the terms of the contributing modelling centres and are not redistributed in this repository. The public package provides the model, member, grid, experiment and variable identifiers, extraction and validation procedures, and compact derived data underlying every figure and table. The version-specific archival record is available on Zenodo at \url{https://doi.org/10.5281/zenodo.21606982}.

\section*{Acknowledgements}
The author acknowledges institutional support from Universidad del Desarrollo. This work received no specific financial support from any public, commercial, or not-for-profit funding agency.

\section*{Author contributions}
M.H.-M. conceived the study, developed the mathematical framework and software, curated and audited the data, performed the analysis, interpreted the results and wrote the manuscript.

\section*{Competing interests}
The author declares no financial or non-financial competing interests.

\end{document}